\documentclass{elsart}
\begin{document}
\begin{frontmatter}


\title{A qualitative Langevin-like  model for the coexistence of two
distinct granular temperatures} 
\author[Rio]{Welles A. M. Morgado\thanksref{FAPERJCNPq}}
\address[Rio]{Departamento de F\'{\i}sica, Pontif\'{\i}cia Universidade
Cat\'olica do Rio de Janeiro \\ CP 38071, 22452-970 Rio de Janeiro,
Brazil\\welles@fis.puc-rio.br}
\thanks[FAPERJCNPq]{Partially supported by FAPERJ and CNPq, Brazil.}

\begin{abstract}
In the present work, we study qualitatively the physics of granular
temperature coexistence, for a mixture of two different species. Our
model captures its essential aspects and this allows us to get insights
on the physical mechanisms of distinct temperature
coexistence, in a way which is not obscured by the
complexities of kinetic theories or numerical simulations.  Our simple
model is consistent with limit situations where we should expect
equality for the granular temperatures for the mixture.
\end{abstract}
\end{frontmatter}


%
%
\section{Introduction}
Granular Systems (GS) are ubiquitous in Nature~\cite{jaeger96}. They
consist of large numbers of distinct, inelastic, rough grains. When
studying GS, we are faced with mesoscopic collective
effects, such as frictional forces, that are not present in molecular
gases~\cite{jaeger96,SOC,shinbrot1}. The presence of energy
dissipation during internal collisions~\cite{brilliantov86} make GS
very different from molecular gases: true equilibrium is only possible
when the total kinetic energy is completely dissipated. Thus, in
order to keep a GS in a steady-state, we need to inject energy through
the boundaries or by means of a coupled heath
bath~\cite{kudrolli1,menon1,wamm_erm,BMPV}.  The parameter used to
account for the kinetic energy present in a GS is the so called
granular temperature, i.e., the average kinetic energy per grain. It
is clearly related to the usual microscopic definition of temperature
in statistical mechanics~\cite{chapman}.

Thus, we define macroscopic granular quantities taking inspiration
from thermodynamic concepts familiar to us. For instance, pressure and
granular number density, among others, are widely used when studying
GS~\cite{chapman,many}.  However, the naive application of
thermodynamic concepts to GS can sometimes be misleading. Take
pressure for instance: for molecular gases, as we increase the number
density, at fixed energy feeding rate, the pressure should
increase. However, for GS, the reverse happens for high enough
densities~\cite{wamm_erm}.  This tells us to be careful when applying
thermodynamic concepts, such as the granular temperature $T_g$, to GS.

Thermodynamics is based on a few simple principles, much in the same
way as Euclidean geometry is based on five postulates.  However, we
can build non-Euclidean geometries from the first four postulates and
the negation of the fifth. We ask ourselves: is there a similar hope
for GS thermodynamics?  In this spirit, we start by studying the most
basic law of granular thermodynamics: the zeroth law, or the
temperature equilibrium law. Our model is a very simple one, based on
a Langevin-like approach to the collisions of different granular
species in a mixture.

We study the transient behavior of a binary granular gas mixture
evolving toward a steady-state~\cite{santosdufty}. The mixture
components are smooth, inelastic identical spheres of two type: a
species of mass $M$ and another species of mass $m\leq M$ and same
radius. A convenient Langevin-like collision term is introduced,
aiming to take into account the more complicated aspects of the
physics in a simplified, effective way.  Experimental realizations of
granular temperature coexistence~\cite{feitosa} have shown some
characteristics that can be reproduced by the present model. In our
opinion, the model's main advantage is that it permits us to visualize
some important physical aspects of an actual system in a mathematical
setting which is much simpler then that of more complete
theories~\cite{garzodufty}.

This paper is organized as follows. In Section 2, we study the
behavior of an intruder in a bath of particles. In Section 3, we
analyze the mixture of granular species while we calculate the
temperature ratio in Section 4. We conclude briefly in Section 5.

%
%
\section{Test particle in a granular bath}
We study the behavior of a 1D test particle of mass $M$, velocity $V$,
immersed in granular bath of particles of mass $m=\gamma M$ and same
radius of the test particle~\cite{brey1}. The granular bath interacts
with the test particle through frontal inelastic collisions
characterized by a coefficient of restitution $0<\varepsilon<1$.

We assume that the typical bath particle has a {\bf constant driving
velocity $v_D$} and we suppose that collisions happen with equal
probabilities for the direction of the bath particle's velocity. The
choice above is a reasonable one, since there are always four possible
ways for a bath particle to collide with the test particle (with
velocity $V$): two frontal collisions (with $V$ and $\pm v_D$) and two
rear ones (again with $V$ and $\pm v_D$). For each case, regardless of
whether $\left| V\right|>v_D$ or $\left| V\right|<v_D$, a simple
analysis shows that the next collision will occur either with a $+v_D$
or with a $-v_D$ bath particle.  A further implicit simplifying
hypothesis is that we ignore the dependence of the collision rate on
the relative velocity. This is similar to the so called Maxwell
model~\cite{ben}.  We thus write the momentum conservation and the
inelastic behavior equation as
\begin{eqnarray}
MV\pm mv_D & = & MV'+mv',\\
V'-v' & = & -\varepsilon (V-(\pm v_D)).
\end{eqnarray} 
After a little manipulation, we obtain
\begin{equation}
V'=\left(\frac{1-\gamma\varepsilon}{1+\gamma}\right)V\pm
v_D\left(\frac{\gamma(1+\varepsilon)}{1+\gamma}\right).
\end{equation}
The test particle's change in velocity is thus
\begin{equation}
\Delta
V=V'-V=\beta[-V+\eta
v_D],\mbox{ where }
\beta=\frac{\gamma(1+\varepsilon)}{1+\gamma},\label{deltav}
\end{equation}
and $\eta$ is a random variable with values $\pm 1$ with equal
probability distribution, $<\eta> = 0$, $\eta^2 =1$.

The test particle's collisional history is determined by its sequence
of collisions with the bath particles.  We assign an independent
random variable $\eta_{N-1}$ for the $N$-th collision. Then, we can
write the velocity of the test particle, after the $N-$th collision as
\begin{eqnarray}
 V_N & =&
(1-\beta)^NV_0+\beta v_D
\sum_{i=0}^{N-1}(1-\beta)^{N-1-i}\eta_i,\label{vsolucao}
\end{eqnarray}
where $V_0$ is the initial velocity of the test particle and
\begin{equation}
<\eta_i> = 0, \mbox{ }<\eta_i\eta_j> =  \delta_{i,j},\mbox{ } \eta_i^2 =1.
\end{equation}
It can be seen from above that the memory of the initial velocity is lost
exponentially fast.

Next, we calculate the cases of the test particle being a bath
particle ($\gamma =1$) and that of it being a massive intruder
($\gamma\neq1$)~\cite{brey1,martin}. In these two limiting cases, the
quadratic velocities will be in equilibrium with the bath and their
ratio should reflect the ratio between test particle-bath particle
quadratic ratio. Obviously, we have $v_{bath}\neq v_D$ since we allow
$v_{bath}$ to fluctuate while supposing $v_D$ fixed. Thus, it is
straightforward to show that
\begin{eqnarray}
v^2_{bath}= \lim_{N\rightarrow\infty,\gamma=1}<V_N^2>
&=&v_D^2\frac{1+\varepsilon}{3-\varepsilon},\label{v21}\\ v^2_{test}=
\lim_{N\rightarrow\infty}<V_N^2> &=&
v_D^2\frac{\gamma(1+\varepsilon)}{2+\gamma(1-\varepsilon)}.\label{v22}
\end{eqnarray}
 
From Eq.~\ref{v21}-\ref{v22}, we obtain the granular temperature
ratio in the steady state for a test particle immersed in a granular
bath: 
\begin{equation}
\frac{T_g^{bath}}{T_g^{test}}=\frac{\frac{1}{2}m<v^2_{bath}>}{\frac{1}{2}M<v^2_{test}>}
=\frac{\frac{m}{2}v_D^2\frac{1+\varepsilon}{3-\varepsilon}}{\frac{
M}{2}v_D^2\frac{\gamma(1+\varepsilon)}{2+\gamma(1-\varepsilon)}}=
\frac{2+\gamma(1-\varepsilon)}{3-\varepsilon}.\label{tracer}
\end{equation}

In order to compare the result of Eq.~\ref{tracer} with the literature
on the tracer limit, we see that in the quasi-elastic case,
$\varepsilon\rightarrow1$, we have up to first order in the limit of
the heavy tracer $\gamma\rightarrow0$
\begin{equation}
\frac{T_g^{test}}{T_g^{bath}}\approx
1+\frac{1-\varepsilon}{2}.\label{traceraprox}
\end{equation}
From reference~\cite{garzodufty}, which agrees with
references~\cite{brey1,martin}, we observe that from Eq. 31 and
Fig. 3, the same limit above is obtained. The present model seems to
have better agreement with kinetic theories on this region. The case
of larger $\gamma$ does not agree so well. However, its aim is not to
obtain accurate quantitative results, but to get insights into the
mechanism of granular temperature coexistence.

The cases of: same species ($\gamma=1$) and elastic motion
($\varepsilon=1$) imply that $T_g^{bath}=T_g^{test}$, as expected.

We observe that the above difference in granular temperatures is due
to the factor $\beta$ being smaller for the massive intruder, $\gamma
<1$~\cite{brey1}, leading to a smaller loss of speed during an
inelastic collision giving
\[
(1-\beta_{int})=\left(1-\frac{2\gamma}{1+\gamma}\frac{(1+\varepsilon)}{2}
\right)>(1-\beta_{bath})=\left(1-\frac{(1+\varepsilon)}{2}\right),
\]
(see Eq.~\ref{vsolucao}). This reflects the fact that, due to its
inertia, a massive inelastic intruder will lose comparatively less
energy to a lighter bath than a less massive particle would.

Our next step will be to characterize the mixture of different
species.

%
%
\section{Granular mixture}
We suppose that the granular fluid under study is composed of
homogeneous particles of identical diameters. Similarly to the
previous section, any chosen particle's collisions will occur with a
bath of particles.  But now this bath consists of a mix of two types
of particles with different constant driving velocities, namely $v_D$
for particles $m$ and $V_D$ for particles $M$~\cite{santosdufty} (the
above mentioned particle belongs to one of these families). The
coefficients of restitution are $\varepsilon_1$, $\varepsilon_2$ and
$\varepsilon'$ corresponding to the $mm$, $MM$, $mM$ (or $Mm$)
collision possibilities. In what follows, we use the definitions
\begin{equation} \alpha_1 =\frac{1+\varepsilon_1}{2},\mbox{ }
\alpha_2 =  \frac{1+\varepsilon_2}{2},\mbox{ }
\beta=\frac{\gamma(1+\varepsilon')}{1+\gamma}=\gamma\zeta.
\end{equation}

We need two families of random variables in order to describe the chosen
particle's $N-$th collision of a: a variable $\theta_{m,N}$ or
$\theta_{M,N}$, to describe with what species the collision occurs;
another, $\eta_{m,N}$ or $\eta_{M,N}$, to give the direction of the
bath particle. These variables have the following distributions:
$\theta_{m,N} = 1$ with probability $x$, $\theta_{m,N} = 0$ with
probability $1-x$, $\theta_{M,N} = 1$ with probability $1-x$,
$\theta_{M,N} = 0$ with probability $x$, where
\begin{equation}
x =\frac{\rho_m}{\rho_m+\rho_M},
\end{equation}
and $\rho_m$ and $\rho_M$ are the species number densities; the
$\eta-$variables are distributed with equal probabilities.

According to the distributions above and skipping the algebra, in the
limit $N\rightarrow\infty$ we obtain: $<v_{\infty}>=0$ and
\begin{eqnarray}
<v_{\infty}^2>
&=&\frac{x\alpha_1^2v_D^2+(1-x)\zeta^2V_D^2}{1-(x(1-\alpha_1)^2+(1-x)
(1-\zeta)^2)}.\label{eqm}
\end{eqnarray}

Similarly, we have  $<V_{\infty}>=0$ and
\begin{eqnarray}
<V_{\infty}^2>
&=&\frac{(1-x)\alpha_2^2V_D^2+x\beta^2v_D^2}{1-((1-x)(1-\alpha_2)^2+
x(1-\beta)^2)}.\label{eqM}
\end{eqnarray}

%
%
\section{Temperature ratio: general situation}
As in the case of a test particle, we define the granular temperature
ratio as
\begin{equation}
\frac{T_g^m}{T_g^M}=\frac{\frac{1}{2}m<v_{\infty}^2>}{\frac{1}{2}M
<V_{\infty}^2>}=\gamma\frac{<v_{\infty}^2>}{<V_{\infty}^2>}=\gamma
\frac{v_D^2}{V_D^2},\label{ratiomM}
\end{equation}
where we assumed that the same relationship holds both for the
steady-state velocities and the driving velocities.

Replacing Eq.~\ref{eqm} and Eq.~\ref{eqM} into Eq.~\ref{ratiomM}, and
solving for $\frac{T_g^m}{T_g^M}$, yields
\begin{eqnarray}
\frac{T_g^m}{T_g^M} &=&\gamma\left\{
\frac{[x\alpha_1^2-(1-x)\alpha_2^2A]}{2x\beta^2A}\right.\nonumber\\
&&\left.+\frac{\sqrt{[x\alpha_1^2-(1-x)\alpha_2^2A]^2
+4x(1-x)\beta^2\zeta^2A}}{2x\beta^2A}\right\},
\label{ratio}
\end{eqnarray}
where
\begin{equation}
A=\frac{1-x(1-\alpha_1)^2-(1-x)(1-\zeta)^2}
{1-(1-x)(1-\alpha_2)^2-x(1-\beta)^2}.
\label{A}
\end{equation}
We notice, in Eq.~\ref{ratio}, that the dependence of
$\frac{T_g^m}{T_g^M}$ on $\gamma$ is stronger than on the inelasticity
parameters, as mentioned in reference~\cite{feitosa}. Some important
physical limits are listed below.

\noindent
{\em Elastic limit.}
In this limit, $\varepsilon_1=\varepsilon_2=\varepsilon'=1$, and we have
\begin{equation}
\alpha_1  =\alpha_2 =  1,\mbox{ }
\beta=\gamma\zeta=\frac{2\gamma}{1+\gamma}\Rightarrow 
\frac{T_g^m}{T_g^M}=1,
\end{equation}
for any value of $\gamma$. The elastic equilibrium regime is such that
all granular temperatures are the same, as expected.

\noindent
{\em Same species, same inelasticity.}
In this case we set $\gamma=1$ and $\varepsilon_1=
\varepsilon_2=\varepsilon'\equiv\varepsilon$. We have now
\[
\alpha_1 = \alpha_2 =\beta =\zeta=\frac{1+\varepsilon}{2}.
\]
It is also trivial to check that $\frac{T_g^m}{T_g^M}=1$. But, for
$\gamma\neq 1$ and $\varepsilon_1,\varepsilon_2,\varepsilon'<1$ the
granular temperature ratio is non trivial, assuming values different
from 1.

\noindent
{\em Tracer limit.}  It is straightforward to show that we recover
Eq.~\ref{tracer} when $x\rightarrow 0$~\cite{brey1,martin}:
\begin{equation}
\lim_{x\rightarrow0}\frac{T_g^{bath}}{T_g^{test}}=
\frac{2+\gamma(1-\varepsilon)}{3-\varepsilon}.
\end{equation}

\noindent
{\em Comparison with experiments.}  In a recent work, Feitosa and
Menon~\cite{feitosa} used beads of equal diameters, made from
different materials in order to study the granular temperature
behavior of a granular mix. According to their experimental data, for
a 1/2-1/2 mix of glass-brass: $x=0.5$, $\gamma =0.28$, $\varepsilon_1
=0.83$, $\varepsilon_2=0.61$, $\varepsilon'=0.72$, $\alpha_1=0.92$,
$\alpha_2=0.81$, $\beta=0.38$ and $\zeta=1.34$. We assumed
$\varepsilon'=\frac{\varepsilon_1+\varepsilon_2}{2}$ in order to carry
out the calculations. By looking at the data in
reference~\cite{feitosa}, we estimate $\frac{T_g^m}{T_g^M}\approx0.8$
as the plateau value for large velocities, where the above ratio seems
to converge to.

From our model, and using the data above, we obtain
\[
\frac{T_g^m}{T_g^M}=0.86,
\]
which is not too far from the experimental result above. But we need
to keep in mind that the present model is basically qualitative. It is
not sophisticated enough in order to obtain precise quantitative
agreement with the experiments.

\section{Conclusions}
A simple Langevin-like collisional model, valid for any mass ratio and
inelasticities values, which assumes a bath of grains interacting with
test particles, is used to study the steady-state average kinetic
energy of a mixture of distinct species of grains. The collisions are
modeled by a Langevin-like impulsive term.  The collisional history of
the test particle is obtained and the limiting mean square velocity
calculated. We compare the mean kinetic energy for each species.
There results a difference in coexisting granular temperatures. We
verify that this is due to inertial effects during a collision of
massive and light particles.

\end{document}